\magnification=1200
\hsize=31pc
\vsize=55 truepc
\hfuzz=2pt
\vfuzz=4pt
\pretolerance=5000
\tolerance=5000
\parskip=0pt plus 1pt
\parindent=16pt
\baselineskip=20 truept

\centerline{\bf Gravitational waveforms from inspiralling compact binaries}
\centerline{\bf to second-post-Newtonian order}

\bigskip\bigskip
\centerline{Luc Blanchet$^1$, Bala R. Iyer$^2$,
Clifford M. Will$^3$ and Alan G. Wiseman$^4$}

\bigskip\bigskip
\noindent{$^1$ D\'epartement d'Astrophysique Relativiste et de
Cosmologie (UPR 176 du CNRS),} \par
{Observatoire de Paris, 92195 Meudon Cedex, France}

\noindent{$^2$ Raman Research Institute, Bangalore 560 080, India}

\noindent{$^3$ McDonnell Center for the Space Sciences,
Department of Physics,} \par
{Washington University, St. Louis, Missouri 63130, USA}

\noindent{$^4$ Theoretical Astrophysics 130-33, California Institute of
Technology,} \par
{Pasadena, CA 91125, USA}

\vskip1.5truecm
\noindent {\bf Abstract.}
The two independent ``plus" and ``cross" polarization waveforms associated
with the gravitational waves emitted by inspiralling, non-spinning,
compact binaries are presented, ready for use in the data analysis of
signals received by future laser interferometer gravitational-wave
detectors such as LIGO and VIRGO.  The computation is based on a
recently derived expression of the gravitational field at the
second-post-Newtonian approximation of general relativity beyond the
dominant (Newtonian) quadrupolar field.  The use of these theoretical
waveforms to make measurements of astrophysical parameters and to test
the nature of relativistic gravity is discussed.

\bigskip\bigskip
\noindent{\bf 1. Introduction}

\medskip
Two large-scale laser interferometer detectors of gravitational waves
are now under construction in the U.S.  (LIGO experiment [1]) and in
Europe (VIRGO experiment [2]).  These experiments should detect the
gravitational waves generated by inspiralling compact binary systems at
cosmological distances [3]--[7].  These are systems of two compact
objects (neutron stars or black holes) spiralling very rapidly around
each other in their late stage of evolution immediately preceding the
final coalescence.  The dynamics of such systems is entirely ruled by
the radiation reaction forces due to the emission of the gravitational
radiation itself. Estimates of the number of inspiralling/coalescing
events each year have been found to be quite promising [8]--[11].

The LIGO and VIRGO observations of inspiralling compact binaries should
provide precise measurements of the masses of the objects, possibly of
their spins and possibly, in the case of neutron stars, of their radii
[12]--[19].  The absolute luminosity distance of the binary will be
measured independently of any assumption concerning the nature (and
masses) of the objects [20].  But if the intrinsic masses of the objects
are known, the cosmological red-shift of the host galaxy where the event
took place can be measured.  There is hope to deduce in this way a
measurement of the Hubble parameter from an assumption concerning the
statistical distribution of masses of neutron stars [21], [22].  The
puzzle of the origin of gamma-ray bursts could be solved by comparison
of the times of arrival of the gravitational waves and of gamma ray
bursts [10], [23], [24].  Furthermore new limits on the validity of
alternative theories of gravity, notably scalar-tensor theories [25],
and new tests of general relativity in the strong field regime [26]
should be possible by monitoring very precisely the inspiralling signal.

To prepare for the analysis of such signals in the LIGO and VIRGO detectors
one needs to compute the gravitational radiation field generated by a
system of two point-masses moving on a circular orbit (the relevant case
because the orbit will have been circularized by radiation reaction
forces).  Since inspiralling compact binaries are very relativistic,
this problem is highly nontrivial and represents a challenge to
relativity theorists [5], [7], as its resolution involves carrying out
the calculation to a very high order in terms of a post-Newtonian
expansion (see Will [27] for a review).  The problem can be decomposed
into two different problems, which can be referred to respectively as
the ``wave generation problem" and the ``radiation reaction problem"
(see Damour [28] for a discussion).

The wave generation problem deals with the computation of the
gravitational waveforms generated by the binary (at the leading order
in $1/r$, where $r$ is the distance of the binary) when the orbital
phase and frequency of the binary take some given values $\phi$ and
$\omega$.  This problem involves computing the amplitude of each
harmonic of the wave corresponding to frequencies which are multiples of
the orbital frequency, with the predominant harmonic being at twice the
orbital frequency.

The radiation reaction problem consists of determining the evolution of the
orbital phase $\phi (t)$ itself as a function of time, from which one
deduces the orbital frequency $\omega (t) = d\phi (t)/dt$.  The actual time
variation of $\phi (t)$ is nonlinear because the orbit evolves under
the effects of gravitational radiation reaction forces.  In principle it
should be determined from the knowledge of the radiation reaction forces
acting locally on the orbit.  However these forces are at present not
known with sufficient accuracy (only the relative first post-Newtonian
corrections are known [29], [30]), so in practice the phase evolution is
determined by equating the high-order post-Newtonian energy flux in the
waves or energy loss (averaged over one orbit) and the decrease of the
correspondingly accurate binding energy of the binary.

Estimates of the precision needed in the resolution of these two
problems can be inferred from black-hole perturbation techniques in the
special case where the mass of one object is very small as compared with
the other mass [31]--[37].  The required precision is reached when the
systematic errors due to the neglect of some higher-order approximation
become less than the statistical errors due to noise in the detector.
It turns out that the post-Newtonian corrections in the time evolution
of the phase (radiation reaction problem) will be measurable in advanced
detectors, probably up to three post-Newtonian (3PN) orders beyond the
(Newtonian) quadrupole radiation [33],[37].  This corresponds to
relativistic corrections in the energy loss as high as order $\sim
(v/c_0)^{6}$ where $v$ is the orbital velocity and where we denote for
later convenience the speed of light by $c_0$.  The possibility of
measuring such high-order corrections can be understood crudely from the
fact that, in order not to suffer a too severe reduction in
signal-to-noise ratio, one will have to monitor the phase evolution with
an accuracy of one tenth of a cycle over the tens of thousands of cycles
during the entire passage through the frequency bandwidth of the
detector.

It has been argued [15]--[17] that most of the accessible information
allowing accurate measurements of the binary's intrinsic parameters
(such as the two masses) is in fact contained in the phase, because of
the accumulation of cycles, and that rather less accurate information is
available in the wave amplitude itself.  For instance, the relative
precision in the determination of the distance $r$ of the source, which
affects the wave amplitude, will be less than for the masses,
which strongly affect the phase evolution.  Consequently, the
determination of $r$ does not necessitate as high a post-Newtonian
precision as for the masses, and to a good approximation it should be
adequate for that purpose to include in the gravitational waveform (as
determined by solving the generation problem) post-Newtonian corrections
up to an order less than 3PN order.  To determine just what order is
required in the waveform would necessitate a measurement-accuracy
analysis similar to the ones performed in Refs.  [17]--[19] but not
restrained to a model waveform taking only the higher-order phase
evolution into account (namely the so-called restricted post-Newtonian
waveform).  To our knowledge no complete investigation has been
performed yet, however we believe that such a study would be desirable.
For instance, it is conceivable that conclusions reached regarding the
needed accuracy in the determination of the phase in the reaction
problem are modified when the full amplitude evolution of the wave is
taken into account.

Both the gravitational waveform and associated outgoing flux or energy loss
have been obtained recently to the 2PN approximation corresponding to  the
relative post-Newtonian order $\sim (v/c_0)^4$. Two computations have been
performed, one by Blanchet, Damour and Iyer [38] using a mixed post-Newtonian
and post-Minkowskian formalism [39], [40], and one by Will and Wiseman
[41] using a purely post-Newtonian formalism due to Epstein and Wagoner
[42] and Thorne [43]. The result of these computations for the energy loss
has been summarized in [44].  More recently, the precision on the energy
loss has been extended to include the next 2.5PN approximation [45].

Even though the 2PN or even 2.5PN precision in the resolution of the
reaction problem appears to be still insufficient to make full use of the
phase data [33], [37], it is plausible, as previously argued
[15]--[17], that the 2PN waveform amplitude is already close to what
will be needed by the LIGO and VIRGO detectors.  However this 2PN
waveform is displayed in Refs.~[38], [41] in a format which is not ready
for use in the future analysis of the outputs of LIGO and VIRGO.  The
aim of this paper is to present the complete, ready-to-use expression of
the waveform, including the amplitude of all the wave harmonics present,
to 2PN order, for non-spinning bodies.  More precisely, we compute the two
independent polarization states of the gravitational wave (customarily
referred to as the ``plus" and ``cross" polarizations) which define the
theoretical ``templates" to be cross correlated with the raw output of
the detectors.  The templates should be used both ``on line" when
searching for the signal and later when the signal is subject to a very
precise data analysis involving a more accurate determination of the
parameters, and possibly the measurement of other parameters.

In Section 2 we present our main result, namely the two plus and cross
polarization waveforms to 2PN order as functions of the phase and
frequency.  For completeness and discussion thereafter we recall also
the results obtained in [38], [41] concerning the time evolution to 2PN order
of the phase and frequency.  In Section 3 we discuss
various issues associated with the actual use of these waveforms.

\bigskip
\noindent {\bf 2. The gravitational waveforms to second-post-Newtonian order}

\medskip
The gravitational wave as it propagates through the detector in the source's
wave zone is entirely described by the transverse and traceless ($\rm
TT$) asymptotic waveform $h_{ij}^{\rm TT} = (g_{ij} - \delta_{ij})^{\rm
TT}$ where $g_{ij}$ denotes the spatial covariant metric in a coordinate
system adapted to the wave zone, and $\delta_{ij}$ is the spatial part
of the Minkowski metric (signature $-+++$).  The two polarization states
$h_{+}$ and $h_{\times}$ are defined by $h_{+} = {1\over 2} (p_i p_j -
q_i q_j) h_{ ij}^{\rm TT}$ and $h_{\times} = {1\over 2} (p_iq_j + p_j
q_i) h_{ij}^{\rm TT}$ where ${\bf p}$ and ${\bf q}$ denote two
polarization vectors forming, along with the direction ${\bf n}$ from the
source to the detector, an orthonormal right-handed triad.  The detector
is directly sensitive to that linear combination of polarization
waveforms $h_+$ and $h_\times$ which is given by
$$
   h (t) = F_{+} h_{+} (t) + F_{\times} h_{\times}(t) \ , \eqno(1)
$$
where $F_{+}$ and $F_{\times}$ are the so-called beam-pattern functions of the
detector depending on two angles giving the direction $-{\bf n}$ of the source
as seen from the detector and a polarization angle specifying the
orientation of the vectors ${\bf p}$ and ${\bf q}$ around that
direction.  The expressions of $F_+$ and $F_\times$ in terms of these
angles are given explicitly in the case of laser-interferometer
detectors by e.g.  Eq.~(104) of Thorne [3].

The two polarizations $h_{+}$ and $h_{\times}$ in the case of a binary made
of non-spinning point-masses moving on a (quasi-)circular orbit are
obtained by a straightforward computation starting from the end results
of Refs.~[38], [41].  We choose the polarization vectors ${\bf p}$ and
${\bf q}$ to lie along the major and minor axis (respectively) of the
projection onto the plane of the sky of the circular orbit, with ${\bf
p}$ oriented toward the ``ascending node'', the point at which body 1
crosses the plane of the sky moving toward the observer.  The ${\rm
TT}$ waveform $h_{ij}^{\rm TT}$ is split as in Eq.~(4.1) of Ref.~[38] into
the sum of an ``instantaneous" contribution $(h_{ij}^{\rm TT})_{\rm inst}$
and of a ``tail" contribution $(h_{ij} ^{\rm TT})_{\rm tail}$ (we comment
below on the presence of wave tails in the signal and their possible
detection).  The instantaneous contribution is given e.g.  by Eqs.~(4.5)
of Ref.~[38] from which we compute first $(h_{+} )_{\rm inst}$ and
$(h_{\times} )_{\rm inst}$.  Then we add to these the tail parts $(h_{+,
\times})_{\rm tail}$ which are already obtained in Eqs.~(4.9)-(4.10) of
Ref.~[38].  Identical formulae result from calculating $h_+$ and
$h_\times$ from the circular-orbit limit of Eqs. (6.10) and (6.11) of
Ref. [41].  Some transformations are necessary in order to express the
result in the best form for applications in the LIGO and VIRGO
detectors.  The final result reads
$$ h_{+,\times} = {2Gm\eta\over c^2_0 r}
   \left({Gm\omega \over c_0^3}\right)^{2/3}
   \left\{ H^{(0)}_{+,\times} + x^{1/2} H^{(1/2)}_{+,\times}
   + x H^{(1)}_{+,\times}
   + x^{3/2} H^{(3/2)}_{+,\times} + x^2 H^{(2)}_{+,\times}
    \right\}\ , \eqno(2)
$$
where the brackets involve a post-Newtonian expansion whose various
post-Newtonian terms are given for the plus polarization by
$$ \eqalignno{
 H^{(0)}_+ &= -(1+c^2) \cos 2\psi \ , &(3a) \cr
 H^{(1/2)}_+ &= -{s\over 8} {{\delta m} \over m} \biggl[ (5+c^2) \cos \psi - 9
  (1+c^2) \cos 3\psi \biggr]\ , &(3b) \cr
 H^{(1)}_+ &= {1 \over 6} \biggl[ ( 19 + 9 c^2 - 2 c^4)
   - \eta ( 19 - 11 c^2 - 6 c^4 ) \biggr] \cos 2\psi     \cr
   &-{4\over 3} s^2 (1+c^2) (1-3\eta) \cos 4\psi\ , &(3c) \cr
 H^{(3/2)}_+ &= {s \over 192}{{\delta m}\over m}\biggl\{\biggl[ (57 + 60 c^2
  - c^4) - 2 \eta (49 - 12 c^2 - c^4) \biggr] \cos \psi  \cr
  &- {27 \over 2} \biggl[ ( 73 + 40 c^2 - 9 c^4)
  - 2 \eta (25 - 8 c^2 - 9 c^4) \biggr] \cos 3\psi \cr
  & + {625\over 2} (1-2\eta) s^2 (1+c^2)\cos 5\psi \biggr\}
   - 2 \pi (1+c^2) \cos 2\psi \ , &(3d) \cr
 H^{(2)}_+ &= {1 \over 120} \biggl[ ( 22 + 396 c^2 + 145 c^4
  - 5 c^6) + {5 \over 3} \eta ( 706 - 216 c^2 - 251 c^4 + 15 c^6)  \cr
 &\qquad\quad -5 \eta^2 (98 - 108 c^2 + 7 c^4 + 5 c^6) \biggr] \cos 2 \psi \cr
 &+ {2 \over 15} s^2 \biggl[ (59 + 35 c^2 - 8 c^4) - {5 \over 3} \eta ( 131
     + 59 c^2 - 24 c^4)  \cr
 &\qquad\quad + 5 \eta^2 (21 - 3 c^2 - 8 c^4) \biggr]  \cos 4 \psi \cr
 &- {81\over 40} (1-5\eta +5\eta^2) s^4 (1+ c^2) \cos 6\psi \cr
 &+ {s \over 40} {{\delta m} \over m}\biggl\{ \biggl[ 11 + 7 c^2 + 10 (5+c^2)
   \ln 2 \biggr] \sin \psi - {5\pi} (5+c^2) \cos \psi \cr
 & \qquad\quad - 27 \biggl[ 7 - 10\, \ln (3/2) \biggr] (1+c^2) \sin 3\psi
   + 135 \pi (1+c^2) \cos 3\psi \biggr\}\ , &(3e) \cr }
$$
and for the cross polarization by
$$ \eqalignno{
 H^{(0)}_\times &= -2c \sin 2\psi\ , &(4a)\cr
 H^{(1/2)}_\times &= - {3\over 4} s c {{\delta m} \over m} \biggl[ \sin \psi
    - 3\sin 3\psi \biggr] \ ,&(4b)\cr
 H^{(1)}_\times &= {c \over 3} \biggl[ ( 17 - 4 c^2) -\eta (13 - 12 c^2)
    \biggr] \sin 2\psi \cr
  &-{8\over 3} (1-3\eta) c s^2 \sin 4\psi\ , &(4c) \cr
 H^{(3/2)}_\times &= {{s c} \over 96} {{\delta m} \over m} \biggl\{ \biggl[ (63
    - 5 c^2) - 2 \eta ( 23 - 5 c^2) \biggr] \sin \psi \cr
 &-{27\over 2}\biggl[ (67 - 15 c^2) -2\eta ( 19 - 15 c^2)\biggr] \sin 3\psi\cr
 &+{625\over 2} (1-2\eta) s^2 \sin 5\psi \biggr\}
   - 4 \pi c \sin 2\psi\ , &(4d) \cr
 H^{(2)}_\times &= {c \over 60} \biggl[ ( 68 + 226 c^2 - 15 c^4) +{5 \over 3}
     \eta (  572 - 490 c^2 + 45 c^4) \cr
 &\qquad - 5 \eta^2 (56 - 70 c^2 +15 c^4 )\biggr] \sin 2\psi \cr
 &+{4 \over 15} cs^2 \biggl[ (55 -12 c^2) -{5 \over 3}\eta ( 119 - 36 c^2)
  + 5 \eta^2 (17 - 12 c^2) \biggr] \sin 4\psi \cr
 & - {81\over 20} (1-5\eta +5\eta^2) c s^4 \sin 6\psi \cr
 &-{3\over 20} sc {{\delta m} \over m} \biggl\{ \biggl[ 3 + 10 \ln 2 \biggr]
     \cos \psi +5 \pi \sin \psi  \cr
 &\qquad\qquad -9 \biggl[ 7 - 10 \ln (3/2) \biggr] \cos 3\psi
  - 45 \pi \sin 3 \psi \biggr\}\ . &(4e) \cr }
$$
The notation is as follows.  The post-Newtonian expansion in (2) is
parametrized by $x\equiv (Gm\omega /c^3_0)^{2/3}$ where $\omega$ is the
2PN-accurate orbital frequency of the circular orbit ($\omega =2\pi/P$
where $P$ is the orbital period) and $m\equiv m_1+m_2$ is the total mass
of the binary.  [The expansion (2) is valid up to the neglect of 2.5PN
terms of order $O(x^{5/2})$.] In addition to $m$ we denote $\delta m
\equiv m_1-m_2$ and $\eta \equiv m_1 m_2/m^2$.  The vector ${\bf n}$
along the line of sight from the binary to the detector defines the
inclination angle $i$ with respect to the normal to the orbital plane.
The normal is chosen to be right-handed with respect to the sense of
motion so that we have $0 \leq i \leq \pi$.  The notations $c$ and $s$
are shorthand for the cosine and sine of the inclination angle~:
$c\equiv \cos i$ and $s\equiv \sin i$.  Finally the basic phase variable
$\psi$ entering (3)-(4) is defined by
$$
 \psi =\phi - {2Gm \omega \over c_0^3} \ln \left ({\omega \over \omega_0}
\right) \ , \eqno(5)
$$
where $\phi$ is the actual orbital phase of the binary, namely the angle
oriented in the sense of motion between the ascending node and the
direction of body 1 ($\phi =0 \, {\rm mod} \, 2\pi$ when the two bodies lie
along ${\bf p}$, with body 1 at the ascending node). The logarithmic term
in the definition of $ \psi$ involves a constant frequency scale $\omega_0$
which can be chosen arbitrarily (see below).  This logarithmic phase
modulation was determined in Refs.~[46], [47] and originates physically
from the propagation of tails in the wave zone. Using $\psi$ instead of the
actual phase $\phi$ simplifies somewhat the expression of the waveform,
since it permits collecting in a single block all the logarithmic terms
(to 2PN order).  The variable $\psi$ is also very convenient in
black-hole perturbation theory, where it permits the resolution of the
Teukolsky equation governing the outgoing radiation up to a very high
order in the post-Newtonian expansion [35], [36].

The expressions (2)-(5) solve the generation problem for
inspiralling compact binaries to 2PN order. Up to now, both the plus and
cross polarization waveforms were known for arbitrary masses to 1.5PN
order [47], [48], and in the test mass limit $\eta \to 0$ to 1.5PN
order [32] and 4PN order [36].  Through 1.5PN order equations (3)
and (4) agree with Wiseman [49] (see also [27]) when those
formulae are reduced to the circular-orbit limit.  Equations (3) and (4)
also largely agree with similar formulae given by Poisson [32] and
Wiseman [47], however those formulae contained some typographical
errors (see [48]).  In any case, the final results given in Eqs.  (3)
and (4) supersede the previous results.  We have checked that our
expressions (3) and (4) in the test mass limit $\eta \to 0$ are in
perfect agreement with the truncation to 2PN order of the results given
in Appendix B of Tagoshi and Sasaki [36]. (The comparison shows that
the phase variable used by Tagoshi and Sasaki is related to ours by
$\psi_{\rm TS} = \psi + (2Gm \omega /c_0^3) [\ln 2 - {17\over 12}]$ which
clearly corresponds simply to a rescaling of the frequency $\omega_0$.)

In order to obtain the expressions of the waveforms $h_+ (t)$ and
$h_\times (t)$ as functions of time one needs to replace $\psi$ and
$\omega$ appearing in equations (2)--(5) by their explicit time
evolutions $\psi (t)$ and $\omega (t)$ obtained from the resolution of
the radiation reaction problem. This problem has been solved in Refs.~[38],
[41] to 2PN order, and we quote here these results for completeness
and later discussion.  It is convenient to introduce instead of the
local time $t$ flowing in the experimenters' reference frame the
dimensionless time variable
$$
\Theta = {c^3_0 \eta \over 5Gm} (t_c - t)\ , \eqno(6)
$$
where $t_c$ is a constant which represents the instant of coalescence
of the two point-masses (at which the frequency goes formally to infinity).
Then the instantaneous orbital phase $\phi (t)$
is given in terms of the time variable (6) by Eq. (4.29) in Ref.~[38] which
reads
$$ \eqalignno{
 \phi (t) &=\phi_c -{1\over\eta} \biggl\{ \Theta^{5/8} +\left({3715\over 8064}
 + {55\over 96}\eta \right) \Theta^{3/8} - {3\pi \over 4} \Theta^{1/4} \cr
&\qquad\qquad\quad\qquad +\left({9275495\over 14450688}+{284875\over 258048}
\eta\ +{1855\over 2048} \eta^2 \right) \Theta^{1/8} \biggr\}\ ,&(7)\cr }
$$
where $\phi_c$ is another constant representing the value of the phase at
$t_c$.
[Note that when taking into account higher order post-Newtonian
approximations (starting at 2.5PN) the phase $\phi (t)$ no longer tends
to a constant when $t \to t_c$ but instead becomes infinite [45].  In
this case the constant $\phi_c$ is simply determined by initial
conditions when the frequency of the wave enters the detector's
bandwidth.]
The orbital frequency is obtained simply by
differentiating equation (7) with respect to time ($\omega = d\phi /dt$),
hence
$$ \eqalignno{
 \omega (t)&= {c_0^3 \over{8Gm}} \biggl\{ \Theta^{-3/8} +\left({743\over 2688}
 + {11\over 32}\eta \right) \Theta^{-5/8} - {3\pi \over 10} \Theta^{-3/4} \cr
 &\qquad\qquad\qquad\qquad + \left( {1855099\over 14450688}
+ {56975\over 258048} \eta + {371\over 2048} \eta^2 \right) \Theta^{-7/8}
\biggr\}\ .&(8) \cr }
$$
 From (7) and (8)  one deduces the phase variable $\psi$ using equation (5).
Both the expressions (7) and (8) are valid up to the 2PN order which
corresponds formally to the same relative precision as for the waveforms
(3) and (4). However it is not a priori required for consistency to
have the same post-Newtonian precision in both the waveforms and phase.
On the contrary, one should use in the waveforms (3) and (4) the best
available expression for $\phi (t)$ which will be hopefully determined
in the future to a much higher order than 2PN (see Ref.~[45] for the
expression of $\phi (t)$ to 2.5PN order).  Related to this, note that
the logarithmic term in the phase variable $\psi (t)$
although of formal order 1.5PN, is actually of order 4PN relatively to the
dominant term in $\phi (t)$ given by (7) (indeed $\phi (t)- \phi_c$ is
of order $\sim c^5_0$ which is the inverse of the order of the
radiation reaction effects).  Thus the logarithmic term is in fact
currently negligible but must be included when the
precision on $\phi (t)$ improves in the future to reach 4PN (even
without knowing $\phi (t)$ to the 4PN order it may be a good idea to
include this term in the templates).

It can be readily checked from the expressions (5)--(8) that any
change in the phase variable $\psi$ corresponding to a rescaling of the
frequency $\omega_0$ is equivalent, to the considered order, to a
shift in the value of the instant of coalescence $t_c$, namely,
the rescaling $\omega_0 \to \lambda\omega_0$ is equivalent to the shift
$t_c \to t_c - (2Gm/c_0^3) \ln \lambda$.  Thus a particular choice of
the frequency scale $\omega_0$ is physically irrelevant since a
different choice can always be absorbed into a redefinition of the
origin of time in the wave zone.
(We have $\omega_0 =(1/4b) \exp ({11\over 12} -C)$ where $C =0.577...$
is Euler's constant and $b$ is a freely specifiable parameter entering
the relation between the wave-zone coordinate time $t$ and the harmonic
coordinates $t_H, r_H$~: $t =t_H- r_H/c_0 -(2Gm/c^3_0) \ln (r_H/c_0b)$
where $r_H$ is the distance of the source in harmonic coordinates, see [46].)
It could be possible to relate $\omega_0$
to the source characteristics, choosing for instance $\omega_0 =
c_0^3/Gm$ where $m$ is the total mass of the binary.  However this
mass will not be known in practice but will be used as a parameter
in the templates to be varied during the data analysis process, so this
choice seems to be somewhat awkward. For practical purposes it is
probably more convenient to choose $\omega_0$ in such a way that it is
uniform over all templates.  For instance one can relate $\omega_0$ to
the detector characteristics by choosing $\omega_0/\pi = 10$ Hz where
$10$ Hz (say) is the seismic cutoff frequency of the detector [26].
Here we adopt this choice.

\bigskip
\noindent{\bf 3. Discussion}

\medskip
To construct adequate filters for the analysis of inspiralling binary signals,
one should proceed as follows. The theoretical waveform $h(t)$ given as
a function of time by the above formulas (1)--(8) is discretized and its
Fourier transform $\tilde h(\Omega)$ is computed numerically and stored. Then
the ratio $\tilde q (\Omega) = \tilde h(\Omega) / S_n (\Omega)$, where $S_n
(\Omega)$ is the measured power spectral density of the noise in the
detector, defines the Fourier transform of the Wiener filter $q(t)$,
which is to be cross-correlated (either in real time
or during the more precise analysis later) with the raw
output $o(t)$ of the detector composed of the superposition of the
actual signal and of the noise (which is supposed here to be Gaussian).
Because of the availability of fast Fourier transforms the correlation is
computed in the Fourier domain using the (discrete) correlation theorem
for the Fourier coefficients $\tilde q (\Omega)$ and $\tilde o(\Omega)$.
This is repeated for each set of parameters in the filter until
maximization of the signal-to-noise ratio is obtained yielding the
determination of the parameters of the binary (if a signal was really
present at this instant).

For two nonrotating test-masses there are 4 parameters entering the phase
of the signal (this is true up to any post-Newtonian order). These can be
chosen to be the two mass parameters $m$ and $\eta$, and the constants
$t_c$ and $\phi_c$. Alternatively one can use the chirp mass ${\cal M}=
\eta^{3/5}m$ and the reduced mass $\mu =\eta m$, and/or the arrival time
$t_0$ and phase $\phi_0$ at the seismic frequency $\omega_0$.  The mass
parameters may also be replaced by two distinct post-Newtonian pieces of
the (chirp) time left till coalescence starting from $\omega_0$ [50].
In addition, the amplitude of each harmonic of the signal depends on
the distance $r$, on the inclination angle $i$, and on the direction of
the binary and the polarization angle through the beam-pattern functions
$F_+$ and $F_\times$. Note that $r$ is the cosmological luminosity
distance, and that the masses are the red-shifted masses which are equal to the
true masses multiplied by $1+z$ where $z$ is the binary's cosmological
red\-shift.  This permits the cosmological measurements proposed in
Refs.~[20]--[22].

In the case of rotating bodies, there are additional parameters in the
signal. Spin-orbit and spin-spin contributions to the waveform and the
phase evolution have been obtained by Kidder {\it et al} [51], [52].
The over-all effect of spins on the accumulated phase during an inspiral
was analyzed by Blanchet {\it et al} [44]. This calculation was done
with the assumption that the spins of the objects were aligned and
perpendicular to the orbital plane.  However, including non-aligned
spins makes the waveform and accumulated-phase calculations considerably
more complicated. In this general case the orbital plane can wobble --
thus the inclination angle $i$ in Eqs.  (3) and (4) changes with time --
giving rise to an amplitude modulation and a frequency modulation
[53].  However note that for realistic inspiralling neutron star binaries
(such as the binary pulsar PSR~1913+16 at the epoch when it finally
coalesces) the spin-orbit
and spin-spin parameters are expected to be very small, so that the dynamics
of these systems will be dominated by the purely gravitational effects
investigated here (see discussion in [44]).

In addition to involving rotating bodies, the binary could move on a slightly
eccentric orbit if, being formed late, it reaches the final inspiral stage
before the circularization of the
orbit by radiation reaction has fully taken place.  This would introduce
an additional term in the phase (7) involving a new parameter which can
conveniently [19] be taken to be $e_0^2 \omega_0^{19/9}$ where $e_0$ is
the initial eccentricity at $\omega_0$ (recall that
$e^2$ evolves in time proportionaly to
$\omega^{-19/9}$ in the quadrupole approximation). The general formulae
of Refs.~[40], [41] and [54] can be
applied to such non-circular situations.

Another possibility is that general relativity is not the
correct theory of gravity because, for instance, of the existence of a
scalar spin-0 field besides the spin-2 metric field.  In this case there
would be still another parameter in the signal describing the coupling of the
scalar field with the matter fields.  Preliminary analyses of the bounds
that could be placed on the coupling parameter $\omega_{\rm BD}$ of the
Brans-Dicke theory can be found in Refs.~[25], [19].

More generally, it should be possible to check that the theoretical waveform
(1)-(8) is exactly reproduced in the real signal without prejudice of
which theory could be the true theory of gravity.  A simple method to do
this consists of parametrizing the templates by a
redundant set of parameters (i.e.  by more parameters than strictly necessary)
and measuring them independently by optimal
filtering.  In this way one can test whether some particular
terms involving some specific combinations of parameters are really
present in the signal [26].  However since
multiplying the number of independent parameters has the effect of
diminishing the accuracy in their measurement, this method necessitates a
high signal-to-noise ratio.

An application of this method is the detection in the real signal of
effects associated with the tails of waves. Physically the tails come
from the backscattering of the linear waves off the static spacetime
curvature generated by the total mass $m$ of the binary itself.  The
tails are characterized by the fact that they are non ``instantaneous'',
namely they depend on the whole integrated past of the source~; however
for binary systems the actual dependence on the past is negligible to
2PN order.  Nevertheless some important signatures of the tail effect
remain in the signal, the most important of which is the
term with coefficient $\pi$ in the phase of the wave (7), which
is directly due to the tails in the radiation reaction forces. This
effect should be easily detectable in the real signal. Other tail
contributions enter the binary's waveform and are due to the propagation
of tails in the wave zone.  An instance is the logarithmic phase
modulation in (5) which can be referred to as a wave tail contribution
to the phase (as opposed to the radiation reaction contribution).  Even
this very small tail-induced phase modulation (of relative 4PN order)
has been shown to be detectable with a high, but not exceedingly high,
signal-to-noise ratio using the above method [26].  This shows how the
observations of inspiralling compact binaries could permit
new tests of general relativity in a regime of strong and
rapidly varying gravitational fields.

\bigskip
\noindent {\bf Acknowledgments}

\medskip
This research is supported in part by the Centre National de la Recherche
Scientifique, by the National Science Foundation
Grants No. PHY 92-22902 (Washington University) and PHY~94-24337
(Caltech) and by the National
Aeronautics and Space Aministration Grants No. NAGW 3874 (Washington
University) and NAGW 4268 (Caltech). One of the authors (A.G.W.) would
like to thank Ben Owen for several helpful comments.

\bigskip
\def\tr#1{\item{#1}}
\noindent {\bf References}

\medskip
\tr{[1]}Abramovici A, Althouse W E, Drever R W P,  G\"ursel Y,
Kawamura S, Raab F J, Shoemaker D, Siewers L, Spero R E,
Thorne K S, Vogt R E, Weiss R, Whitcomb S E and Zucker M E, 1992, Science,
{\bf 256}, 32.

\tr{[2]}Bradaschia C, Calloni E, Cobal M, Del~Fasbro R,
Di~Virgilio A, Giazotto A, Holloway L E, Kautzky H, Michelozzi B,
Montelatici V, Pascuello D and Velloso W, 1991, Proc. of the Banff
Summer Inst. on Gravitation 1990, Banff, Alberta, edited by Mann R and
Wesson P (Singapore:  World Scientific) p. 499.

\tr{[3]}Thorne K S, 1987, 300 Years of Gravitation, edited by
Hawking S W and Israel W (Cambridge: Cambridge Univ. Press) p. 330.

\tr{[4]}Schutz B F, 1989, Class. Quantum Grav., {\bf 6}, 1761.

\tr{[5]}Schutz B F, 1993, Class. Quantum Grav., {\bf 10}, S135.

\tr{[6]}Thorne K S, 1993, Proc.  of the Fourth Rencontres de Blois,
edited by Fontaine G and Tran Thanh Van J (Editions Fronti\`eres, France)

\tr{[7]}Thorne K S, 1994, Proc. of the 8th Nishinomiya-Yukawa
Symposium on Rela\-tivistic Cosmology, edited by Sasaki M (Universal Acad.
Press, Japan) p. 67.

\tr{[8]}Clark J P A, van den Heuvel E P J and
Sutantyo W, 1979, Astron. Astrophys., {\bf 72}, 120.

\tr{[9]}Phinney E S, 1991, Astrophys. J., {\bf 380}, L17.

\tr{[10]}Narayan R, Piran T and Shemi A, 1991, Astrophys.
J., {\bf 379}, L17.

\tr{[11]}Tutukov A V and Yungelson L R, 1993, Mon. Not. R.
Astron. Soc., {\bf 260}, 675.

\tr{[12]}Kr\'olak A and Schutz B F, 1987, Gen. Rel. Grav., {\bf 19}, 1163.

\tr{[13]}Kr\'olak A, 1989, Proc. in Gravitational Wave Data
Analysis, edited by Schutz B F, (Kluwer Academic Publishers).

\tr{[14]}Lincoln C W and Will C M, 1990, Phys. Rev. D{\bf 42}, 1123.

\tr{[15]}Cutler C,  Apostolatos T A, Bildsten L, Finn L S,
Flanagan E E, Kennefick D, Markovic D M, Ori A, Poisson E,
Sussman G J and Thorne K S, 1993, Phys. Rev. Lett., {\bf 70}, 2984.

\tr{[16]}Finn L S and Chernoff D F, 1993, Phys. Rev. D{\bf 47}, 2198.

\tr{[17]}Cutler C and Flanagan E, 1994, Phys. Rev. D{\bf 49}, 2658.

\tr{[18]}Poisson E and Will C M, 1995, Phys. Rev. D{\bf 52}, 848.

\tr{[19]}Kr\'olak A, Kokkotas K D and Sch\"afer G, 1995, Phys. Rev.
D{\bf 52}, 2089.

\tr{[20]}Schutz B F, 1986, Nature, {\bf 323}, 310.

\tr{[21]}Markovi\'c D, 1993, Phys. Rev. D{\bf 48}, 4738.

\tr{[22]}Finn L S, 1995, Proc. of the 17th Texas Symposium
on Relativistic Astrophysics, edited by B\"ohringer H et al (Annals of the
NY Acad. Sc.) p. 489.

\tr{[23]}Paczynsky B, 1991, Acta Astron., {\bf 41}, 157.

\tr{[24]}Eichler D {\it et al}, 1989, Nature, {\bf 340}, 126.

\tr{[25]}Will C M, 1994, Phys. Rev. D{\bf 50}, 6058.

\tr{[26]}Blanchet L and Sathyaprakash B S, 1994, Class. Quantum
Grav., {\bf 11}, 2807 ; and  1995, Phys. Rev. Letters, {\bf 74}, 1067.

\tr{[27]}Will C M, 1994, Proc. of the 8th Nishinomiya-Yukawa
Symposium on Relativistic Cosmology, edited by Sasaki M (Universal Acad.
Press, Japan) p. 83.

\tr{[28]}Damour T, 1986, in Gravitation in Astrophysics, edited by
Carter B and Hartle J B (Plenum Press, New York and London) p~3.

\tr{[29]}Iyer B R and Will C M, 1993, Phys. Rev. Lett., {\bf 70}, 13 ;
and 1995, Phys. Rev. D{\bf 52}, 6882.

\tr{[30]}Blanchet L, 1993, Phys. Rev. D{\bf 47}, 4392~; and
Phys. Rev. D, (submitted).

\tr{[31]}Gal'tsov D V, Matiukhin A A and Petukhov V I, 1980,
Phys. Lett., {\bf 77A}, 387.

\tr{[32]}Poisson E, 1993, Phys. Rev. D{\bf 47}, 1497.

\tr{[33]}Cutler C, Finn L S, Poisson E and  Sussmann G J, 1993,
Phys. Rev. D{\bf 47}, 1511.

\tr{[34]}Tagoshi H and Nakamura T, 1994, Phys. Rev. D{\bf 49}, 4016.

\tr{[35]}Sasaki M, 1994, Prog. Theor. Phys.,{\bf 92}, 17.

\tr{[36]}Tagoshi H and Sasaki M, 1994, Prog. Theor. Phys., {\bf 92}, 745.

\tr{[37]}Poisson E, 1995, Phys. Rev. D{\bf 52}, 5719.

\tr{[38]}Blanchet L, Damour T and Iyer B R,  1995,
Phys. Rev. D{\bf 51}, 5360.

\tr{[39]}Blanchet L and Damour T, 1989,  Ann. Inst. H. Poincar\'e
(Phys. Th\'eorique), {\bf 50}, 377 ;
Damour T and Iyer B R, 1991,  Ann. Inst. H. Poincar\'e
(Phys.  Th\'eorique) {\bf 54} 115 ;
Blanchet L and Damour T, 1992,  Phys.  Rev.  D{\bf 46}, 4304.

\tr{[40]}Blanchet L, 1995, Phys. Rev. D{\bf 51}, 2559.

\tr{[41]}Will C M and Wiseman A G, 1995, in preparation.

\tr{[42]}Epstein R and Wagoner R V, 1975, Astrophys. J., {\bf 197}, 717.

\tr{[43]}Thorne K S, 1980, Rev. Mod. Phys., {\bf 52}, 299.

\tr{[44]}Blanchet L, Damour T, Iyer B R, Will C M
and Wiseman A G, 1995, Phys. Rev. Lett., {\bf 74}, 3515.

\tr{[45]}Blanchet L, Phys. Rev. D, (submitted).

\tr{[46]}Blanchet L and Sch\"afer G, 1993, Class. Quantum Grav., {\bf 10}, 2699.

\tr{[47]}Wiseman A G, 1993, Phys. Rev. D{\bf 48}, 4757.

\tr{[48]}Alexandre J, 1994, Rayonnement gravitationnel \'emis par un syst\`eme
binaire en coalescence, (unpublished).

\tr{[49]}Wiseman A G, 1992, Phys. Rev. D{\bf 46}, 1517.

\tr{[50]}Balasubramanian R, Sathyaprakash B S and Dhurandhar
S V, Phys. Rev. D, (submitted).

\tr{[51]}Kidder L E, Will C M and Wiseman A G, 1993, Phys.
Rev. D{\bf 47}, R4183.

\tr{[52]}Kidder L E, 1995, Phys. Rev. D{\bf 52}, 821.

\tr{[53]}Apostolatos T A, Cutler C, Sussman G J and Thorne K S, 1994,
Phys. Rev. D{\bf 49}, 6274.

\tr{[54]}Gopakumar A and Iyer B R, 1995, in preparation.
\bye